\documentclass{aa}  
\usepackage{hyperref}
\usepackage{amsmath}
\usepackage{graphicx}
\usepackage{tabularx}
\usepackage{multirow}
\usepackage{txfonts}
\newcommand{\Ang}{\text{\AA}}
\begin{document} 

   \title{A new code for low-resolution spectral identification of white dwarf binary candidates}
 
   \subtitle{ }

   \author{Genghao Liu \inst{1,2} 
        \and Baitian Tang \inst{1,2}\thanks{\email{tangbt@mail.sysu.edu.cn}}
        \and Liangliang Ren \inst{3}
        \and Chengyuan Li \inst{1,2}
        \and Sihao Cheng \inst{4}
        \and Weikai Zong \inst{5,6}
        \and Jianning Fu \inst{5,6}
        \and Bo Ma \inst{1,2}
        \and Cheng Xu \inst{1,2}
        \and Yiming Hu \inst{1,7}
          }
   \institute{School of Physics and Astronomy, Sun Yat-sen University, Zhuhai 519082, China
         \and CSST Science Center for the Guangdong-Hong Kong-Macau Greater Bay Area, Zhuhai 519082, China
         \and School of Electrical and Electronic Engineering, Anhui Science and Technology University, Bengbu, Anhui 233030, China
         \and School of Natural Sciences, Institute for Advanced Study, Princeton, 1 Einstein Drive, NJ 08540, USA
         \and Institute for Frontiers in Astronomy and Astrophysics, Beijing Normal University, Beijing 102206, China
         \and School of Physics and Astronomy, Beijing Normal University, Beijing 100875, China
         \and MOE Key Laboratory of TianQin Mission, TianQin Research Center for Gravitational Physics, Frontiers Science Center for TianQin, Gravitational Wave Research Center of CNSA, Sun Yat-sen University, Zhuhai 519082, China
             }

   \date{Received 08/07, 2024; accepted 31/07, 2024}

  \abstract
   {Close white dwarf binaries (CWDBs) are considered to be progenitors of several exotic astronomical phenomena (e.g., type Ia supernovae, cataclysmic variables). These violent events are broadly used in studies of general relativity and cosmology. However, obtaining precise stellar parameter measurements for both components of CWDBs is a challenging task given their low luminosities, swift time variation, and complex orbits. High-resolution spectra (R$> 20 000$) are preferred but expensive, resulting in a sample size that is insufficient for robust population study. Recently, studies have shown that the more accessible low-resolution (R$\sim 2000$) spectra (LRS) may also provide enough information for spectral decomposition. To release the full potential of the less expensive low-resolution spectroscopic surveys, and thus greatly expand the CWDB sample size, it is necessary to develop a robust pipeline for spectra decomposition and analysis.}
   { We aim to develop a spectroscopic fitting program for white dwarf binary systems based on  photometry, LRS, and stellar evolutionary models. The outputs include stellar parameters of both companions in the binary including effective temperature, surface gravity, mass, radius, and metallicity in the case of MS stars.}
   { 
   We used an artificial neural network (ANN) to build spectrum generators for DA/DB white dwarfs and main-sequence stars. Characteristic spectral lines were used to decompose the spectrum of each component. The best-fit stellar parameters were obtained by finding the least $\chi^2$ solution to these feature lines and the continuum simultaneously. Compared to previous studies, our code is innovative in the following aspects: (1) implementing a sophisticated binary decomposition technique in LRS for the first time; (2) using flux-calibrated spectra instead of photometry plus spectral lines, in which the latter requires multi-epoch observations; (3) applying an ANN in binary decomposition, which significantly improves the efficiency and accuracy of generated spectra.}
   {
   We demonstrate the reliability of our code with two well-studied CWDBs, WD 1534+503 and PG 1224+309. We also estimate the stellar parameters of 14 newly identified CWDB candidates, most of which are fitted with double component models for the first time. Our estimates agree with previous results for the common stars and follow the statistical distribution in the literature. 
   }
   {We provide a robust program for fitting binary spectra of various resolutions. Its application to a large volume of white dwarf binary candidates will offer important statistic samples to stellar evolution studies and future gravitational wave monitoring.}


   \maketitle
%

\section{Introduction}
\label{sec:intro}

Close white dwarf binaries (CWDBs), consisting of at least one white dwarf in the system, are possible descendants of common envelope evolution \citep[e.g.,][]{Gokhale2007,anguiano_white_2022}. They are categorized as one type of post-common envelope binary \citep{Gokhale2007}. Due to the complex mass loss mechanism and short timescale, 
the progenitors of CWDBs — common envelope binaries (CE) — lack well-measured stellar parameters, and thus lead to insufficient theoretical interpretation. The observational constraints from post-common envelope binaries greatly illuminate the studies of CE \citep{zhan-wen_han_binary_2020}. On the other hand, the detection of CWDBs may also identify the progenitors of many exotic objects, such as over-luminous type IA supernovae. 

When both components of CWDBs are white dwarfs, they are named double white dwarfs (DWDs), which are the main source of galactic gravitational wave foreground radiation in the frequency ranging from $10^{-4}$ to $0.1$ Hz. A number of different population models predict that $\sim 10^{4}$ of close DWDs (0.1\% of Galactic ultra-compact binaries) will be resolved by the Laser Interferometer Space Antenna \citep[LISA, ][]{georgousi_gravitational_2022,amaroseoane2017laser}. But until the year 2022, only 25 DWDs had well-measured orbital parameters and mass ratios \citep{Kilic_2020}, and only 150 DWDs had known orbital parameters \citep{korol_observationally_2022}. 

Many CWDB candidates have been discovered by time-domain photometric surveys, such as the Kepler K2 mission \citep{howell_k2_2014,hallakoun_sdss_2016,van_sluijs_occurrence_2018} and the Zwicky Transient Facility \citep[ZTF; see ][]{keller_eclipsing_2021,burdge_systematic_2020,burdge_88_2020,burdge_general_2019}.  \cite{Ren_2023} used photometric data from Gaia and ZTF to identify a group of CWDB candidates. Their orbital periods and classifications were determined by fitting light curves. Their results provide valuable samples for further spectroscopic study. In the near future, the upcoming Chinese Space Station Telescope (CSST) will detect faint sources down to $\sim26-27$ mag in multiple passbands (ultraviolet to optical), greatly expanding the samples of CWDB candidates. 

Follow-up confirmation of CWDB candidates generally relies on high-resolution spectra (HRS, $R \gtrsim 20,000$). In this case, the double lines from both components are mostly resolved, and thus  precise stellar parameters and chemical abundances can be measured. For example, \cite{Bédard_2017} demonstrated that it is possible to derive atmospheric parameters of DA+DA  binaries by combining spectroscopic (R$\sim 18500$), photometric, and astrometric measurements.\footnote{Also see updated parallax measurements from \cite{Bergeron_2019}.} Their work broke the parameter degeneracies by fitting Balmer lines and spectral energy distributions (SEDs) simultaneously. They identified 15 double degenerate binary candidates out of 219 CWDBs. In a follow-up study, \cite{Kilic_2020} performed multi-epoch high resolution (R$\sim 18000$ for the Gemini Multi-Object Spectrograph, $R \sim 37000$ for the Keck I HIRES echelle spectrograph) spectral analysis on four double degenerate binary candidates. They first determined orbital parameters of each binary with radial velocity (RV) measurements, and then fit the Balmer lines plus the SED to obtain effective temperatures and surface gravities. Finally, the white dwarf (WD) masses were determined through evolutionary tracks.

However, the high-resolution spectrum is so expensive that the resulting sample size is generally small.   
To fully embrace the arrival of large surveys with a lower spectral resolution, such as the Sloan Digital Sky Survey V \citep[SDSS-V, ][]{kollmeier2019sdss} and the Large Sky Area Multi-Object Fiber Spectroscopic Telescope \citep[LAMOST, ][]{Zhao_2012}, a few attempts have been made. It is challenging to decompose the flux contributions from both components since the line-blending effect in low-resolution spectra (LRS) is severe. 
The parameter degeneracy is attenuated in CWDBs with different spectral type components, because of the significant SED discrepancy between the two components. For example, the white dwarf + dwarf M (WD+dM) binary can be decomposed by fitting the red and blue ends of the spectrum individually, as WDs dominate the flux at the blue end, while M dwarfs dominate the red end. For example, \cite{Ren_2013} utilized principal component analysis and template-matching techniques 
 to obtain stellar parameters for ten WD+dM binaries based on LAMOST LRS. 

During the process of estimating the stellar parameters of binary systems, two strategies are usually adopted to generate the model synthetic spectra: template-matching (e.g., \citealt{Ren_2013}) and real-time computing (e.g., iSPEC, \citealt{2019MNRAS.486.2075B}). Matching spectra with a pre-computed template grid limits the precision of the template resolution, while resolving the radiation transfer equation in real time is time-consuming. A trade-off method is to apply a machine-learning method or polynomial interpolation to parameterize the synthetic spectra. With these techniques, synthetic spectra can be generated faster and their parameter coverage is more continuous. Artificial neural network (ANN) parameterization is a popular algorithm, which employs codes such as \texttt{PAYNE}  \citep{ting2019payne,xiang2022stellar} and \texttt{wdtools} \citep{chandra2020computational}. The ANN behaves better than 21-term polynomial interpolation \citep[e.g., ][TGM2 in \texttt{Ulyss}]{sharma_stellar_2020}.

Compared to HRS, LRS generally have the advantages of a wider wavelength range, a higher signal-to-noise ratio (S/N), and a shorter exposure time. A short exposure time is particularly important when measuring the RV of close binaries with short periods (of the order of minutes to hours), given that long exposure times can smear the RV signal completely \citep{rebassa-mansergas_where_2019}. The information loss due to the line-blending effect is compensated for by a wider wavelength range coverage and higher S/N. If we have robust models and the ability to fit all stellar labels simultaneously, the uncertainty should be entirely independent of the resolution. This holds over spectral resolutions from $R \sim 1000$ to $R \sim 100,000$ \citep{ting_prospects_2017}. At the same time, the large number of unresolved binary candidates with LRS call for a robust pipeline to determine the parameters of both components.

This work aims to develop a pipeline to derive the stellar parameters of both components of CWDBs based on LRS. The inputs of our code include: flux-calibrated spectra and parallax. The outputs include: the atmospheric parameters ($\mathrm{T_{\mathrm{eff}},logg}$ for WDs, $\mathrm{T_{\mathrm{eff}},logg}$, and $\mathrm{[Fe/H]}$ for MS), mass, and radius of each component. Our code is innovative compared to previous studies; for example, (1) implementing the binary decomposition technique of \cite{Kilic_2020} in LRS for the first time, (2) using the flux-calibrated spectra, instead of SED plus spectra lines (the latter requires multi-epoch observations), and (3) applying an ANN in binary decomposition for the first time, significantly improving the efficiency and accuracy of generated spectra. 
We structure this paper as follows. In Sect. \ref{cha:data}, we outline the procedures of our code and show how to generate model spectra using an ANN. In Sect. \ref{cha:sp}, we show how to determine stellar parameters by simultaneously fitting line profiles and continuum. After testing the credibility of our code, we apply it to the SDSS LRS of 14 WDB candidates in Sect. \ref{VandA}. In Sect. \ref{discussion}, we discuss our results and compare them to previous works. In Sect. \ref{conclusion}, we briefly summarize this work and give a future perspective.


\section{Generating model spectra}

\label{cha:data}

The procedures of our code are the following:
(1) For a given CWDB, we first evaluated the possible stellar combinations according to its location in the color magnitude diagram (CMD) \citep[see figures of][]{inight_towards_2021}. 
(2) We generated synthetic spectra of both stellar components based on known spectral libraries. In this work, we used the ANN stellar spectrum generator.
(3) We then fit the flux-calibrated spectra to derive the atmospheric parameters of both components.
(4) We finally estimated the mass and radius by interpolating between evolutionary tracks.
Given that our code assumes a combination of two stellar components, it is not suitable to fit binary systems with significant emission lines, which indicate prominent binary interaction.
We first describe how to generate model spectra using the ANN stellar spectrum generator in this section.\footnote{We do not include subdwarfs in this work, since their overlapping evolutionary tracks \citep[e.g.,][]{han2002origin} could cause severe problem when inferring the mass and radius (Sect. \ref{subsec:massandra}). }

\subsection{Spectral grids}

Main-sequence stars: The Coelho synthetic stellar library \citep{2014MNRAS.440.1027C} is a high-resolution (constant wavelength sampling of $\Delta \lambda = 0.02\Ang$) theoretical spectral database for main-sequence stars. This library covers B-type to M-type main-sequence stars in the following parameter space:
$$2500\Ang \leq \lambda \leq 9000\Ang$$
$$3000 \mathrm{K} \leq T_{\mathrm{eff}} \leq 25000 \mathrm{K}$$
$$4.0 \leq \mathrm{log g} \leq 5.5 $$
$$-1.3 \leq \mathrm{[Fe/H]} \leq 0.2$$
$$\mathrm{[\alpha/Fe]} = 0.0,0.4.$$

\noindent White dwarfs: Among WDs, there are $\sim 80\%$ DA-type WDs (DAs)\footnote{WDs with an almost pure hydrogen atmosphere, featuring broad Balmer lines.} and $\sim 20 \%$ DB/O-type WDs (DB/O)\footnote{WDs with He atmosphere: DB is featured by He I lines; DO is featured by He II lines.} \citep{sion1983proposed,kx2019}. Thus, we have ignored other type of WDs in this work. Here, we used the grid of pure-H and pure-He atmospheres from \citet{koester2010white}.

The pure-H templates\footnote{The pure-H templates can be downloaded at \href{http://svo2.cab.inta-csic.es/theory/newov2/index.php?models=koester2}{SVO theoretical services}} cover the following parameter space:
$$900\Ang \leq \lambda \leq 30000\Ang$$
$$5000 \mathrm{K} \leq T_{\mathrm{eff}} \leq 80000 \mathrm{K}$$
$$6.5 \leq \mathrm{log g} \leq 9.5 .$$

\noindent The pure-He templates cover the following parameter space:
$$3000\Ang \leq \lambda \leq 9000\Ang$$
$$8000 K\leq T_{\mathrm{eff}} \leq 40000 K$$ 
$$7.0 \leq \mathrm{\log g} \leq 9.5.$$

\subsection{Artificial neural network stellar spectrum generator}
\label{cha:generator}

\begin{figure*}[!h]
    \centering
    \includegraphics[width=\textwidth]{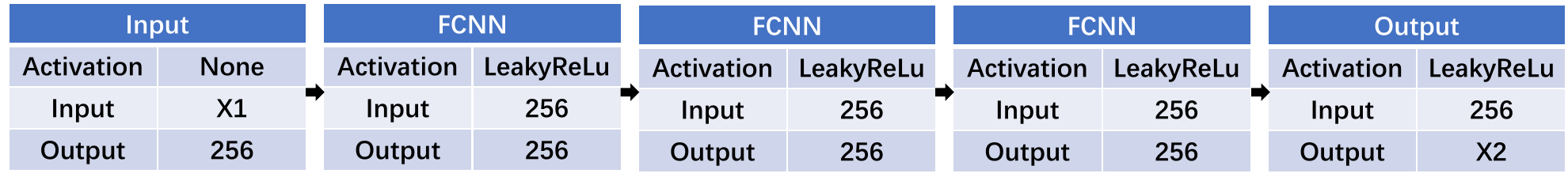}
    \caption{Conceptual illustration of the ANN architecture. 
    X1 is the number of input parameters, while X2 is the number of output pixels.
    }
    \label{frame}
\end{figure*}

Inspired by the work of \citet{ting2019payne}, we adopted an ANN architecture. It is fully connected with three hidden layers, each consisting of 256 neurons (Fig. \ref{frame}). The normalized flux ($f_{\lambda}$), as a function of the stellar parameter label ($l$), can be written as follows:
\begin{equation}
f_\lambda=w^{i}_{\lambda}A\left(w_\lambda^j A\left(w^{k}_{\lambda ij} l_{k}+b_{\lambda ij}\right)+b_{\lambda i}\right)+\Bar{f_{\lambda}},
\end{equation} 
where $\lambda$ represents the index of the output wavelength pixels, and indices $k,j,i$ represent the neuron orders of the first, second, and third ANN layers, respectively. $A$ represents the activation function of a neuron, where a leaky ReLU activation function (Eq. \ref{leaky}) is employed. 
\begin{equation}
\label{leaky}
    A(x)=\left\{ 
    \begin{array}{ll}
    {x} & {x>0}\\
    {\alpha x} & {x\leq 0,\alpha=0.1}
    \end{array}\right.,
\end{equation}
The optimizer seeks the coefficients $(w^{i}_{\lambda},w^{j}_{\lambda i},w^{k}_{\lambda ij},\Bar{f_{\lambda}},b_{\lambda i} ,b_{\lambda ij})$ that best fit the training spectra as a function of labels.

Next, we randomly selected a set of parameters as inputs for the spectrum generator. The resulting  flux-calibrated spectrum was compared with theoretical spectrum in the spectral library with the same parameters. 
Our spectrum generator performs smoothly across most of the parameter space with high accuracy (See Appendix \ref{app:norm} for details).

One of our major improvements is the capability to generate flux-calibrated spectra, which are essential for the following fitting procedures. Previous studies always applied continuum normalization to the synthetic spectra before training \citep[e.g.,][]{chandra2020computational,ting2019payne}, which fails to generate flux-calibrated spectra. The technique of continuum-normalization can be described as
\begin{equation}
    \label{scalen}
    F_{\mathrm{norm}}(\lambda) = \frac{F(\lambda)-F_{\mathrm{min}}}{F_{\mathrm{max}}-F_{\mathrm{min}}},
\end{equation}
where $F_{\mathrm{max}}$ and $F_{\mathrm{min}}$ represent the minimum and maximum flux values of a given spectrum, respectively. Without this normalization, the neural network might struggle to achieve accurate interpolation results for some stellar parameters. For example, surface gravity mainly affects the line widths, which have little contribution to the mean square error (MSE) loss function in the neural network. However, the generated spectra after such flux normalization are also scaled, which is not suitable for our code. 

We used a new normalization formula instead of Eq. \ref{scalen}:
\begin{equation}
    \label{scale}
    F_{\mathrm{norm, new}}(\lambda) = \frac{F(\lambda)-F_{\lambda,\mathrm{min}}}{F_{\lambda,\mathrm{max}}-F_{\lambda,\mathrm{min}}},
\end{equation}
where $F_{\lambda,\mathrm{min}}$ ($F_{\lambda,\mathrm{max}}$) represents the minimum (the maximum) flux value among the entire spectral template sets at wavelength $\lambda$.

The flux values of any single parameter set share the same scaling factor ($F_{\lambda,\mathrm{min}}$, $F_{\lambda,\mathrm{max}}$) at wavelength $\lambda$. This allows us to save the “minimum” and “maximum” scaling spectra, $F_{\lambda,\mathrm{min}}$ and $F_{\lambda,\mathrm{max}}$. We can restore the flux-calibrated spectra by reversing the scaling.

\section{Determining stellar parameters}
\label{cha:sp}
\subsection{Spectral preparation before fitting}
\label{sub:prep}

In this work, we intend to use LRS as the input, but the blended line profiles are insufficient to decompose both components. Inspired by the work of \citet{Kilic_2020}, we split the LRS into line segments and continuum segments, and prepared them separately before fitting. We note that the continuum segments are defined as the spectral regions outside of the line segments. In total, 234 frequently used feature lines (of 38 types of atoms or ions) were selected from the National Institute of Standards and Technology (NIST) Atomic Spectra Database.\footnote{\href{https://physics.nist.gov/PhysRefData/Handbook/Tables/findinglist1.html}{https://physics.nist.gov/}}

The RV, caused by the Doppler effect or the gravitational effect, should be determined prior to fitting stellar parameters, given that the RV changes the observed wavelengths of the lines. The most persistent feature lines, such as Na D for M stars, Fe I for F-K stars, $H\alpha $ for A stars, and He for O-B stars \citep{hilditch2001introduction}, were used to determine the RVs. 

We prepared the line segments and the continuum segments separately. 
For the former, we used the normalized spectra to maximize the changes in the line profiles. For the latter,  we used flux-calibrated spectra. The model spectra can be expressed as

\begin{equation}
\label{line}
        F_{\text{line}}(\lambda) = \frac{R_{1}^{2}S_{1}(T,logg,\lambda)+R_{2}^{2}S_{2}(T,logg,\lambda)}{D^{2}P(\lambda)},
\end{equation}
\begin{equation}
\label{cont}
        F_{\text{cont}}(\lambda) = \left(\frac{R_{1}}{D}\right)^{2}S_{1}(T,logg,\lambda)+\left(\frac{R_{2}}{D}\right)^{2}S_{2}(T,logg,\lambda),
\end{equation}
where $R_{1}$ and $R_{2}$ are the radii determined by the evolution tracks of each stellar component (i.e., $\mathrm{R=R(M(T,\log g),\log g)}$, see Sec.\ref{subsec:massandra} for details), $D$ denotes the distance, $S_{1},S_{2}$ refer to the synthetic spectra of both components, and $P(\lambda)$ represents the pseudo-continuum. Given that the distance, $D$, can be derived from trigonometric parallax measurements, the right-hand sides of these equations only depend on the stellar parameters of both components. The observational spectra were also divided into the corresponding continuum and normalized line segments:
\begin{equation}
    F_{\text{line,observed}}(\lambda) = \frac{O(\lambda)}{P(\lambda)},  F_{\text{cont,observed}}(\lambda) = O(\lambda)
,\end{equation}
where $O(\lambda)$ are the observed spectra and $P(\lambda)$ is the pseudo-continuum.

\subsection{Atmospheric parameters}
\label{sec:fit}
A first guess of the parameters is preferred before the fitting, since it makes the later optimization convergence faster. \cite{Zhang_2019} statistically confirmed the hypothesis that the temperatures of binary spectra derived by \texttt{Ulyss}\footnote{\texttt{Ulyss} parameterizes the 2000 spectra in the ELODIE spectral library for each wavelength pixel in the $T_{\mathrm{eff}}$ range using a 21-term polynomial of $T_{\mathrm{eff}}$, $\mathrm{[Fe/H]}$, $\mathrm{log g}$ in the range from $3900 $ ~\AA~ to $6800 $ ~\AA.} are approximately equal to those of the luminosity-dominant component. Otherwise, if their flux contributions are similar (i.e., not dominated by any components), they probably have a similar temperature. In addition, the metallicities of both components are thought to be equal if they come from the same cloud. 
 
The atmospheric parameters were obtained by simultaneously fitting the line segments and the continuum segments.  We used the “minimize” optimizer from the SciPy package in this work. 
Due to the discrepancy in pixel numbers and flux values between $F_{\text{line}}(\lambda)$ and $F_{\text{cont}}(\lambda)$, one cannot simply add up the residuals of the two segments. We solved this problem by adjusting the weights of the residuals of $F_{\text{cont}}(\lambda)$ and $F_{\text{line}}(\lambda)$. Generally, we decreased the weight of the $F_{\text{cont}}(\lambda)$, so that the line segments and the continuum segments could be fitted equally well. The predicted spectrum is expressed as $Y_{predict}=\boldsymbol{X} L$, where $L$ represents the input parameters and $X$ are the spectral models. 
The matrix of the squared residuals can be written as

\begin{equation}
    \boldsymbol{R}(L) = \frac{1}{2}\left(Y_{\mathrm{predict}}-\boldsymbol{Y}\right)^T\left(Y_{\mathrm{predict}}-\boldsymbol{Y}\right)=\frac{1}{2}\left(\boldsymbol{X}L-\boldsymbol{Y}\right)^T\left(\boldsymbol{X}L-\boldsymbol{Y}\right).
\end{equation}
where Y is the observed spectrum. To represent the weights of different segments, we introduced a transformation matrix, $\boldsymbol{\alpha}$, to the residual:

\begin{equation}
   (\boldsymbol{X} L-\boldsymbol{Y}) \xrightarrow{} \alpha(\boldsymbol{X} L-\boldsymbol{Y}).
\end{equation}
The new squared residual, $\boldsymbol{\Tilde{R}}(L)$, is expressed as

\begin{equation}
    \boldsymbol{\Tilde{R}}(L) = \frac{1}{2}\left(\boldsymbol{X}L-\boldsymbol{Y}\right)^T\boldsymbol{\alpha}^{T}\boldsymbol{\alpha}\left(\boldsymbol{X}L-\boldsymbol{Y}\right).
\end{equation}
The minimal squared residual is located where its gradient equals zero. The gradient is written as
\begin{equation}
\label{jp}
\begin{aligned}
    \frac{\partial}{\partial L} \boldsymbol{\Tilde{R}}(L) &=\boldsymbol{X}^T\boldsymbol{\alpha}^{T}\boldsymbol{\alpha}(\boldsymbol{X} L-\boldsymbol{Y}) \\
    &=\boldsymbol{X}^T\boldsymbol{A}(\boldsymbol{X} L-\boldsymbol{Y})
\end{aligned},
\end{equation}
where each element of the transformation matrix, $\boldsymbol{A}=\boldsymbol{\alpha}^{T}\boldsymbol{\alpha}$, can be regarded as the weight of each wavelength pixel. The transformation matrix, $\boldsymbol{\alpha}$, is set to be

\begin{equation}
    \boldsymbol{\alpha} = 
    \begin{pmatrix}
\frac{1}{F_{\text{med}}} & 0 & 0 & \cdots & 0 & 0 \\
0 & \frac{1}{F_{\text{med}}} & 0 & \cdots & 0 & 0 \\
\vdots & & \ddots & & & \\
0 & \cdots & 0 & 1 & \cdots & 0 \\
0 & \cdots & 0 & 0 & \ddots & 0 \\
0 & \cdots & \cdots & \cdots & 0 & \frac{1}{F_{\text{med}}}
\end{pmatrix}.
\end{equation}
This is a diagonal matrix: elements corresponding to the continuum segments were set to $1/F_{\text{med}}$ ($F_{\text{med}}$ is the median flux of the continuum segments), while elements corresponding to the line segments were set to 1. This transformation reduces the weights of the continuum segment residuals, leading to more reasonable fitting results. 

While looking for the minimal squared residual, Eq. \ref{jp} indicates that the model, $\boldsymbol{X}$, plays an important role. The spectral model, $\boldsymbol{X}$, affects the predicted feature line strength (part of spectrum $Y_{\mathrm{predict}}$) at a given  parameter, $L$. For example, Fig. 1 of \citet{xiang2019abundance} shows the derivatives of stellar spectra with respect to 24 parameters. In other words, some feature lines may be particularly strong (or weak) in one type of stars, which helps to break the degeneracy. In that sense, binaries with significantly different atmospheric parameters are expected to be better decomposed (and vice versa; see Sect. \ref{VandA} for fitting results).

\subsection{Stellar mass and radius}
\label{subsec:massandra}

Assuming that the evolutionary tracks do not overlap, we can determine stellar mass on a three-dimensional grid ($\mathrm{log(T_{\mathrm{eff}}),log g,[Fe/H]}$). Additionally, stellar mass is used to determine the radius of the star and then constrain the model flux through Eq.\ref{line}. Therefore, the stellar radius is not a free parameter in our code, reducing the chance of non-physical results from Markov Chain Monte Carlo fitting.
In this work, we utilized the linear N-dimensional interpolator package (\texttt{LinearNDInterpolator}).\footnote{\href{https://www.osgeo.cn/scipy/reference/generated/scipy.interpolate.LinearNDInterpolator.html}{scipy.interpolate.LinearNDInterpolator.html}} 
The white dwarf evolutionary tracks were obtained from the Python package \texttt{WD\_models} \citep{Cheng_2020}.\footnote{\href{https://github.com/SihaoCheng/WD\_models}{https://github.com/SihaoCheng/WD\_models} base on cooling models from \cite{bedard2020spectral} and atmosphere color table computed and calibrated by \cite{holberg2006calibration,tremblay2011improved,blouin2018new,bergeron2011comprehensive,kowalski2006found} }  The MS stellar evolutionary tracks were picked from MIST,\footnote{\href{http://waps.cfa.harvard.edu/MIST/model_grids.html\#eeps}{http://waps.cfa.harvard.edu/MIST/model\_grids.html\#eeps}} which covers the following parameter space:
$$0.1 \leq \mathrm{M_{initial}/M_{\odot}} \leq 10$$
$$-1.5 \leq \mathrm{[Fe/H]} \leq 0.5$$
$$\mathrm{[\alpha/Fe]} = 0$$
$$\mathrm{V/V_{crit}}=0,0.4,$$
where $\mathrm{M_{initial}}$ is the initial stellar mass and $\mathrm{V_{crit}}$ is the critical linear velocity at the surface\citep{choiMESAIsochronesStellar2016}.

We confirmed the reliability of the interpolation function by reproducing the mass-luminosity and mass-radius relations \citep{2012StellarAstrophysics}.
Next, we calculated the mass uncertainty numerically using the gradients of the evolutionary tracks, and the radius uncertainty was derived through uncertainty propagation.

\section{Verification and results}
\label{VandA}

Our code can be applied to LRS with various resolutions. In this work, we chose SEGUE LRS from the Sloan Digital Sky Survey \citep[][SDSS]{2006AJ....131.2332G}. These spectra cover a wavelength range of $3800$ \AA~- 9200 \AA~with a spectral resolution of $R \sim 1800-2200$.

\subsection{Test sample}
\label{refsp}
To validate our code, we selected two CWDBs with accurate stellar parameter estimation as the reference sample (Table \ref{testsample}). They show distinct double lines in HRS; moreover, time-domain measurements, such as light curves and RVs, are available for these sources. The test sample size is small as the aforementioned high-quality datasets are challenging to obtain.

\begin{table*}[!h]
    \centering
    \caption{Properties of test samples.}
    \label{testsample}
\begin{tabular}{c|c|c|c|c}
    \hline \hline
    &WD 1534+503&This Work& PG 1224+309&This Work\\
    &\citet{10.1093/mnras/stab439}& &\citet{orosz1999post} \\\hline
    $T_{1} (\text{K})$     &$8900^{+500}_{-500} $            & $9085^{+531}_{-725}$      &  $29300$            & $30996^{+217}_{-239}$\\
    $\log g_{1}$&$7.6^{+0.15}_{-0.15}$              &$7.50^{+0.024}_{-0.019}$                & $7.38$                  &$7.43^{+0.009}_{-0.014} $      \\
    $M_{1} (M_{\odot})$     &$0.392^{+0.069}_{-0.059}$  & $0.351^{+0.010}_{-0.007}  $     &  $0.4\pm 0.05$& $0.41^{+0.004}_{-0.004}$                    \\
    $R_{1} (10^{-2}R_{\odot})$     &$1.64$                  & $1.75^{+0.02}_{-0.02} $   & /                              & $ 2.00^{+0.013}_{-0.012}$                      \\
    \hline
    $T_{2} (\text{K})$     &$8500^{+500}_{-500}$            & $7987^{+134}_{-138}$    &  $3100(fixed)$   & $3484.77^{+120}_{-110}$        \\
    $\log g_{2}$&$8.03^{+0.18}_{-0.1}$              & $8.50^{+0.057}_{-0.053}$             &  /                             & $4.99^{+0.004}_{-0.017}$        \\
    $\mathrm{[Fe/H]}$      &/                                           &/                 &  /                             &$-0.047^{+0.11}_{-0.007}$           \\
    $M_{2}(M_{\odot})$&$0.617^{+0.110}_{-0.096} $& $0.917^{+0.035}_{-0.032}$       &  $0.28\pm 0.05 $& $0.40^{+0.005}_{-0.042}$                         \\
    $R_{2}(10^{-2}R_{\odot})$&$1.2567 $                & $ 0.8912^{+0.04}_{-0.05}$& /                              & $23.8^{+0.25}_{-1.64}$                    \\
    \hline\hline
\end{tabular}
\tablefoot{The “WD 1534+503” and “PG 1224+309” columns show the literature fitting results using high-resolution spectra, while our fitting results are listed under the columns ``This work''. The label “/” means no value.}
\end{table*}

\begin{figure*}
    \centering
    \includegraphics[width=\textwidth]{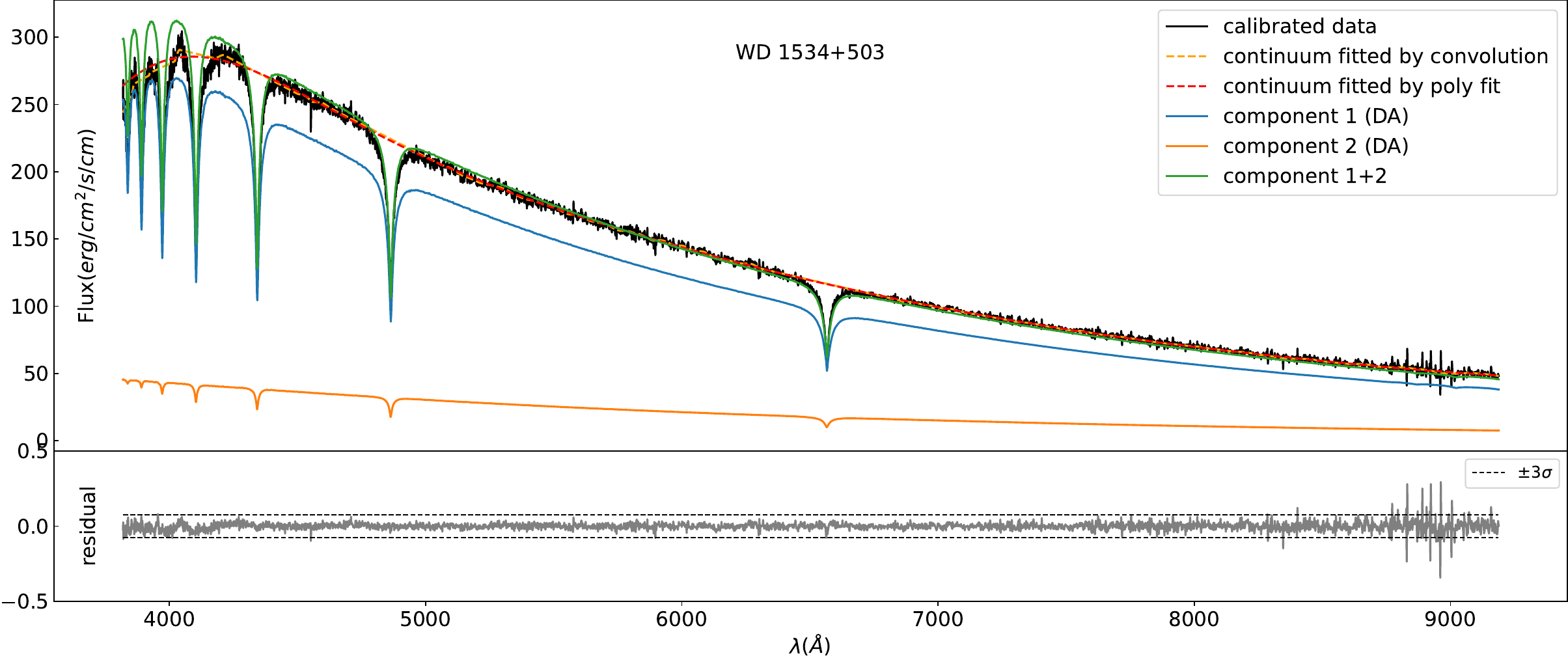}
    \caption{Fitting results for WD 1534+503. The top panel shows the SDSS spectrum (black line), the convolved smoothed continuum (dashed orange line), the pseudo-continuum obtained by polynomial fitting (dashed red line), the best-fit spectrum of Component 1 (blue line), the best-fit spectrum of Component 2 (solid orange line), and the composite spectrum (green line). The bottom panel shows the residual spectrum. The $\pm 3\sigma$ levels are indicated by the dashed lines.}
    \label{sampledwds}
\end{figure*}

\begin{figure}
    \centering
    \includegraphics[width=.5\textwidth]{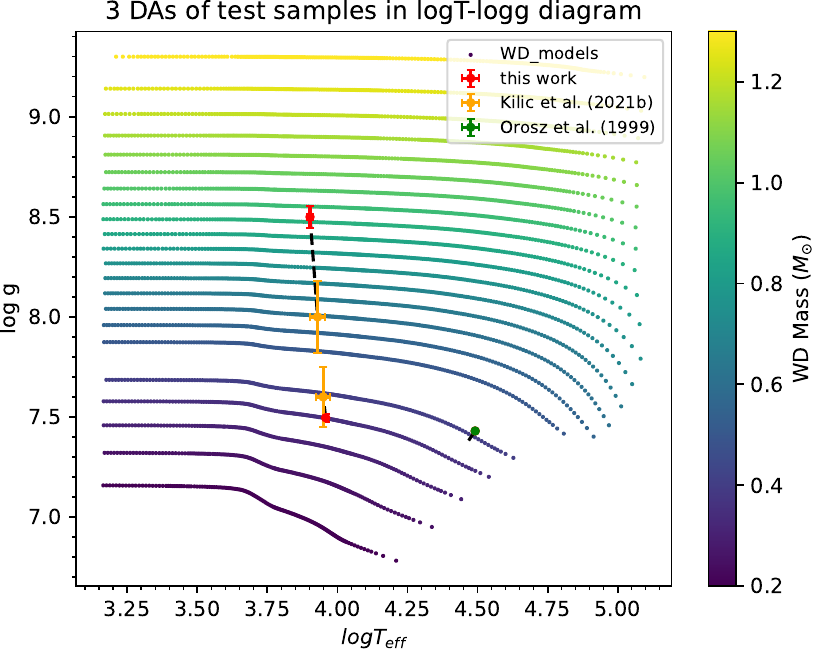}
    \caption{Kiel diagram of three DAs in the test sample. The evolutionary tracks provided by \cite{Cheng_2020} are shown as small dots, where their masses are indicated by the color bar. The red symbols with error bars represent the expected values and uncertainties of our fitting results. Similarly, orange symbols and green symbols show the literature results of \citet{10.1093/mnras/stab439}  and \citet{orosz1999post}, respectively. Obviously, the error in logg is propagated to the stellar mass, and thus the radius. As the surface gravity is expressed on a log scale, and the mass and radius are expressed on a linear scale, the mass-radius errors are magnified. }
    \label{masserr}
\end{figure}

The first CWDB, WD 1534+503, was identified as a binary system with DWDs. \citet{10.1093/mnras/stab439} obtained the orbital and atmospheric parameters of WD 1534+503 with high confidence (Table \ref{testsample}). In this work, our prediction successfully reproduce the flux-calibrated LRS, given that most of the residuals are within the $3\sigma$ level. In particular, the Balmer lines are distinguishable in the residual spectrum, indicating an accurate fitting (the bottom panel of Fig. \ref{sampledwds}). The fitted parameters of the primary star and the fitted effective temperature of the secondary star are consistent with the literature, while the surface gravity of the secondary star is not (see Fig. \ref{sampledwds}). The latter also leads to a significant mass-radius discrepancy (see Fig. \ref{masserr}). 

\begin{figure*}
    \centering
    \includegraphics[width=\textwidth]{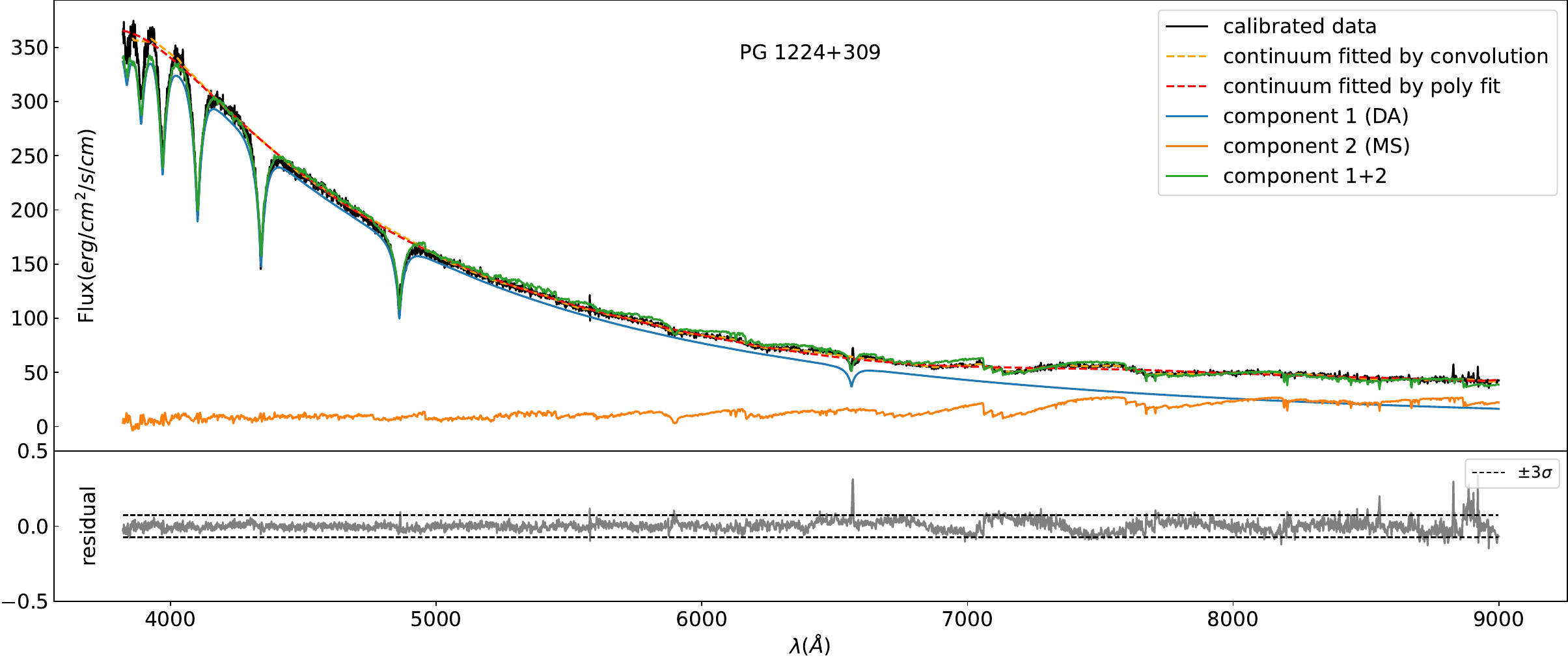}
    \caption{Fitting results for PG 1224+309. Symbols are the same as Fig. \ref{sampledwds}.}
    \label{sampledwdms}
\end{figure*}
The second CWDB, PG 1224+309, is a DA+dM4 binary according to \cite{orosz1999post}. In their study, the masses of both components were obtained from the orbital parameters of this binary system. Given the low luminosity of the M dwarf, its temperature was fixed to 3100 K during the fitting (Table \ref{testsample}).

In this work, our prediction generally reproduces the flux-calibrated LRS, since most of the residuals are within the $3\sigma$ level. We also notice two features in the residual spectrum: (1) distinct features near $7000-8000$ \AA, probably due to the inaccurate modeling of the molecular lines in M dwarfs; (2) a strong $H\alpha$ emission line, which is also visible in the observed spectrum. This is not surprising, given that PG 1224+309 is a short-period eclipsing binary. These defects may be the reason for the inconsistency in the stellar parameters of the secondary star (Table \ref{testsample}).

Through these two CWDBs with high-quality datasets, we prove that our code works reasonably well for LRS with $R\sim 2000$: (1) the stellar parameters of the primary stars are accurate; (2) given the lower luminosities of the secondary stars, their derived atmospheric parameters may be slightly less accurate, leading to larger errors in the estimated masses and radii.

\subsection{Results of 14 white dwarf binary candidates}
\label{source}

After estimating the reliability of our code on LRS, we plan to apply it to photometric identified CWDB candidates with LRS observations. As was mentioned in Sect. \ref{sec:intro}, \cite{Ren_2023} have provided a catalog of 423 short-period eclipsing binary candidates. Among them, 28 sources have SDSS LRS. We further excluded: (1) spectra with S/N below 10; (2) emission-line dominated spectra; (3) previously identified subdwarf$+$M binaries  (J162256.66+473051.1, J011339.09+225738.9, and J082053.53+000843.4, \citealt{Dai_2022,Devarapalli_2022}). We treated the remaining 14 sources as our final sample (Table \ref{var1}).

\begin{figure*}[h]
    \centering
    \includegraphics[width=0.9\textwidth]{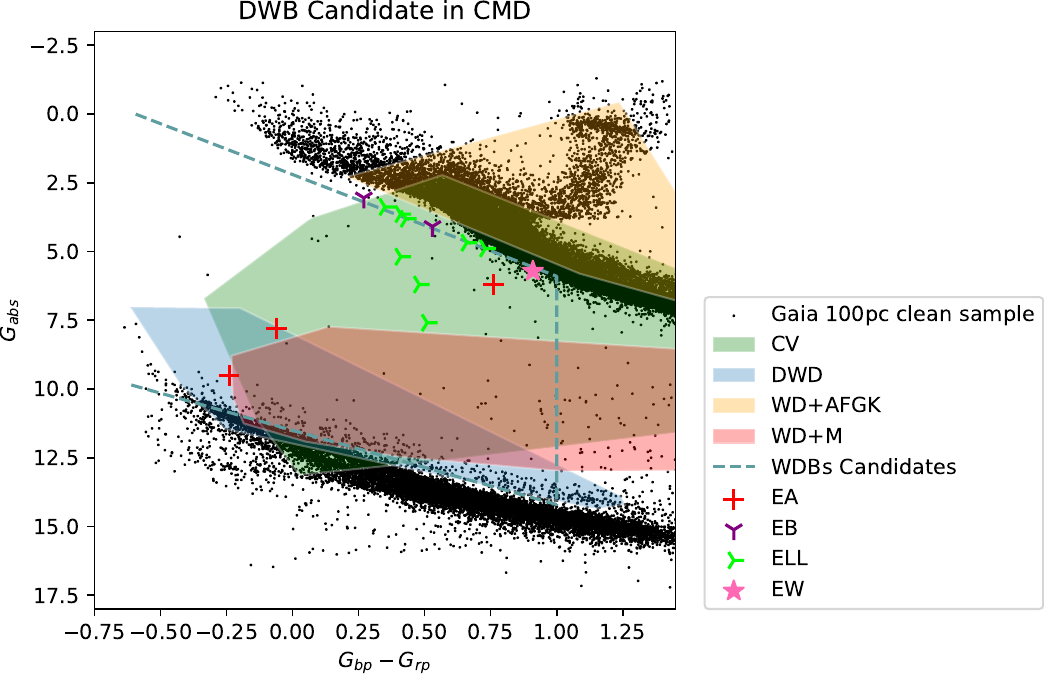}
    \caption{Our final samples on the Gaia CMD (colored symbols). Different types of binary systems identified by \citet{Ren_2023} are denoted by different colored  symbols. The black dots denote Gaia samples within 100 pc with small errors. The region outlined by cadet dashed blue lines was set by \citet{Ren_2023} to select their candidates. The color-shaded regions indicate the locations of gold samples (CVs-green, DWDs-blue, WD+AFGK-orange, WD+M-red) compiled by \cite{inight_towards_2021}.}
    \label{ca}
\end{figure*}

As was mentioned in Sect. \ref{sec:fit}, we first estimated the possible stellar combinations based on their CMD locations (Fig. \ref{ca}). Our sample sources are located between the MS and WD cooling sequence. Based on their colors, the possible primary stars should be MSs or WDs earlier than G-type stars \citet{Zhang_2019}, while the secondary stars have no strong constraint on their spectral type. We further reduced the searching parameter space using the results of \cite{inight_towards_2021}, which show the possible CMD location of CVs, DWDs, WD+AFGK, and WD+M (Fig. \ref{ca}).

For each possible stellar combination, we generated the spectra for both components, and added up their fluxes according to Eqs. \ref{cont} and \ref{line}. The most probable stellar combination was found by fitting the observed spectra (Eq. \ref{sec:fit}). Though our sources show detectable luminosity variation, the secondary star may not be significant enough to be decomposed from single-epoch LRS (too faint or blocked by the primary star). In this work, we consider a source to be a single star if the fitted temperature of the secondary star shows a large uncertainty ($\sigma_{T} > 30\%$  T). The fitted parameters and spectra of our sample are shown in Table \ref{var1} and Fig. \ref{ff}, respectively.

\section{Discussion}
\label{discussion}

\subsection{Residual spectra}
The residual spectra reveal possible features of  interacting  binaries (accretion disk, Roche lobe overflow, stellar wind, etc.). 
 A few sources exhibit multiple metal or molecular lines around 8000-9000 \AA~(e.g., J034137.68+011027.6, J082145.28+455923.4, J110045.15+521043.8, and J093934.56+531751.5) in the residual spectra. These DAZ-like features suggest an accretion of tidally disrupted asteroids and planets\citep{bedardSpectralEvolutionWhite2024}.

\subsection{Statistics}

Though it has a limited sample size, it is interesting to discuss our fitting results in the context of other WD+MS studies. Our analysis identifies one MS+DB, five DA+MS, one DA, two MSs, and five MS+MS. We note that only six out of 14 CWDB candidates have been confirmed, proving the necessity of LRS follow-up observations. 
Our DA mean mass is 0.509 $M_{\odot}$, which is consistent with the peak value of WD mass distribution in \citet[][]{ren2018white}, while our MS mean mass is 0.77 $M_{\odot}$. We further divided DA+MS into DA+dM and DA+AFGK systems, to compare them with literature studies.
For the two DA+dM binaries in our samples, the mean WD temperature is 54612 K, which is much higher than those reported in two previous studies   \citep[$\sim$15000 K,][]{nayak2023hunting, ren2018white}. This discrepancy is caused by their different colors: the targets in our work are bluer ($G_{bp}-G_{rp} < 1$) compared to the aforementioned studies ($G_{bp}-G_{rp} > 1$). Binaries with hotter primary components tend to have bluer colors. As for the three DA+AFGK binaries in our sample, the mean temperature of DA and MS agrees with the peak value of the DA+AFGK temperature distribution given by \cite[][]{ren_white_2020}.

\subsection{Individual analysis}

ZTF J110045.15+521043.80 is identified as a hot DA WD accompanied by an invisible star. The WD dominates almost all the luminosity. Residual analysis reveals the presence of metal and OH I emission lines in the spectral region of $>8000$ \AA, suggesting a potential accretion of surrounding dust and cool dense molecular clouds. Our fitting results are consistent with those of \cite{kepler2019white}, $T = 23083\pm 208 \mathrm{K}$, $\log g = 7.525\pm 0.028$.

ZTF J140847.17+295044.87is identified as a DA+dM system in this work. 
Our fitting WD temperature ($T_{\mathrm{DA}}=31118 K$) is consistent with the literature ( $T_{\mathrm{DA}} = 30000 K$ in \citealt{morgan2012effects} and $T_{\mathrm{WD}} = 29 050 \pm 484 K$ in \citealt{parsons2013eclipsing}), while the fitting WD surface gravity and mass ($\log g_{\mathrm{DA}} = 8.058$, $M_{\mathrm{DA}} = 0.68 M_{\odot}$) are slightly higher than these studies ($\log g_{\mathrm{DA}} \sim 7.75$, $M_{\mathrm{DA}} \sim 0.5 M_{\odot}$).

Its spectrum displays narrow Balmer emission lines, similar to PG 1224+309 (Sect. \ref{VandA}), indicating that the near side of the MS star is heated by the WD \citep[][]{sing2004spectroscopic}. 

ZTF J082145.29+455923.46 is classified as a DA+MS system in this work. Our classification is consistent with  \cite{rebassa2010post} and  \cite{silvestri2007new}. 

A single WD spectroscopic fitting suggests that $ T = 30384\pm 152 \mathrm{K}$ and $\log g = 7.980\pm 0.034$ \citep{kepler2019white}. On the other hand, \cite{parsons2013eclipsing} classified this source as a WD+M2 binary system using light curve fitting. Their derived WD temperature is  $T_{\mathrm{WD}} = 80 938 \pm 4024 K$, with $M_{\mathrm{WD}} = 0.66 \pm 0.05 M_{\odot}$, which are consistent 
with our prediction.  The discrepancy between our measurements and those of the former study is possibly caused by (1) the orbital phase variation or (2) the defect of single component fitting.

ZTF J034137.67+011027.88 is identified as an MS star with an invisible companion in this work. The spectrum lacks prominent Balmer lines, suggesting there is no strong WD contribution. The spectrum shows  metal emission lines (e.g., Ca I, Na D, Ca II, and Fe I) at the red end, indicating a binary accretion. 

ZTF J143257.08+491143.01 is classified as an MS+MS system in this work. Single MS spectroscopic fitting suggests that $T_{\mathrm{eff}} = 8419\pm 31 K$, $\log g = 4.2\pm 0.03$ and $\mathrm{[Fe/H]} = -1.03\pm 0.05 $ \citep{chen2020stellar}, which is consistent with our prediction of the primary star.

ZTF J140118.80-081723.51 is classified as a DA+MS system in this work. We note that this source was identified as a different type of system in other studies: an extremely low mass (ELM) WD \citep{brown2020elm} or an sdA \citep{kepler2019white}. The former study derived $T = 8813\pm 90 K$, $\log g = 5.731 \pm 0.048$ by fitting its spectrum with a grid of pure hydrogen-atmosphere models. The masses of both components were determined by RV measurements: $M_{\mathrm{WD}} = 0.216 \pm 0.042 M_{\odot}$ is $M_{2,\mathrm{min}}\footnote{minimum mass} = 0.79 M_{\odot}$. Our fitted results do not agree with them. The residual spectrum (Fig. \ref{ff}) shows non-negligible Balmer absorption lines, which may be attributed to the lack of the ELM or sub-dwarf models in our work.

The remaining sources have undergone no previous spectroscopic analysis, but some of them are listed in several catalogs \citep{heinze2018first, marsh2017characterization, drake2014catalina, kepler2019white}. Readers are referred to these works for more details.

\vspace{0.2cm}
\noindent To summarize, we find that our code performs reasonably well when comparing to other studies statistically or comparing fitting results individually. Our fitted residual spectra are also promising to study mild binary activities. We again recognize the disadvantage of our code: the lack of stellar models out of binary evolution; for example, involving subdwarfs and ELM WDs.  

\section{Conclusion}
\label{conclusion}

Close white dwarf binaries play a significant role in understanding the evolution of stellar post-common-envelope phases, and they are also progenitors of several gravitational-wave-emitting objects. The follow-up identification of CWDBs often relies on HRS. However, HRS are too expensive and insufficient to satisfy the growing demands. On the other hand, LRS, like those from SDSS and LAMOST, are more accessible. The LRS also have the advantages of a wider wavelength range, higher S/N, and shorter exposure time.
In particular, upcoming photometric surveys (e.g., CSST, LSST, Euclid) will provide a large sample of faint CWDB candidates beyond the magnitude limit of HRS, even with a 10-meter telescope (e.g., Keck, GTC). The most feasible option is degrading the spectral resolution to achieve enough S/N. Therefore, decoupling CWDB components from LRS is a promising task.

In this work, we develop a new code for fitting LRS ($R \sim 2000$) of CWDB candidates. Our code estimates the atmospheric parameters and masses of both components based on flux-calibrated spectra and parallax. 
We first generated the spectra of WDs and MS stars using an ANN. We then obtained the stellar parameters by fitting the feature lines and continuum simultaneously. Our code has several advantages:
(1) We applied the binary decomposition method  of \cite{Kilic2020} in LRS for the first time. (2) We used flux-calibrated spectra, rather than line profile plus SED data for binary decomposition. This not only avoids the phase variation between spectra and SED observations, but also increases the accessibility of the required data.

(3) The ANN spectrum generator has the advantages of the template-matching technique and the synthetic spectra computing technique, which can generate the model spectra both efficiently and accurately.

By comparing to two CWDBs with well-measured components, we have proven that our code performs well when estimating the stellar parameters of the primary stars. The estimation is less accurate for the secondary stars, given their lower luminosities.

We then applied our code to decompose the stellar parameters of 14 white dwarf binary candidates suggested by \citet{Ren_2023}. We find that only six candidates are CWDBs, proving the necessity of LRS spectroscopic decomposition. The fitting results from our code agree with others from the literature statistically. The individual fitting results are generally consistent with other studies. Two caveats are noted with regard to the implementation of our code: (1) our code is not suitable to fit binary systems with significant emission lines; (2) our code lacks stellar models outside of binary evolution, e.g., subdwarf and ELM ones. These should be the goals of future improvements.

To sum up, our code offers a fully automated approach to identify the most probable stellar combinations and atmospheric parameters of both components. It is a valuable tool for conducting pilot spectroscopic study on CWDB candidates rapidly. Given the large number of LRS available  in current and future spectroscopic studies (e.g., LAMOST-II, DESI), our code will provide a large sample of CWDBs for studying stellar evolution and monitoring gravitational waves.

\begin{acknowledgements}
We thank Jingkun Zhao for helpful discussions.
G.L., B.T., and C.L. gratefully acknowledge support from the China Manned Space Project under grant NOs. CMS-CSST-2021-B03, CMS-CSST-2021-A08, etc., the National Natural Science Foundation of China through grants No. 12233013, and the Natural Science Foundation of Guangdong Province under grant No. 2022A1515010732. 
JNF acknowledges the support from the National Natural Science Foundation of China
(NSFC) through the grants 12090040 and 12090042, and the science research grants
from the China Manned Space Project.

This work utilizes data from the Gaia satellite, a space mission of the European Space Agency (ESA). The Gaia data is provided by the Gaia Data Processing and Analysis Consortium (DPAC). The funding for DPAC comes from national agencies, particularly institutions involved in the Gaia Multilateral Agreement.
This work also utilizes filter curve and theoretical spectral model data from various surveys collected and organized by the Spanish Virtual Observatory (SVO). The funding for SVO is provided by the Spanish Ministry of Science and Innovation (Ministerio de Ciencia e Innovación).
The TMAW tool (http://astro.uni-tuebingen.de/~TMAW) used for this paper was constructed as part of the activities of the German Astrophysical Virtual Observatory\citep{10.1093/mnras/sty056}.
\end{acknowledgements}
\bibliographystyle{aa}
\bibliography{aanda}
\begin{appendix}

\onecolumn

\section{Additional comparison}
\label{app:norm}

\begin{figure}[!h]
\begin{minipage}[c]{0.5\textwidth}
    \centering
    \includegraphics[width=\textwidth]{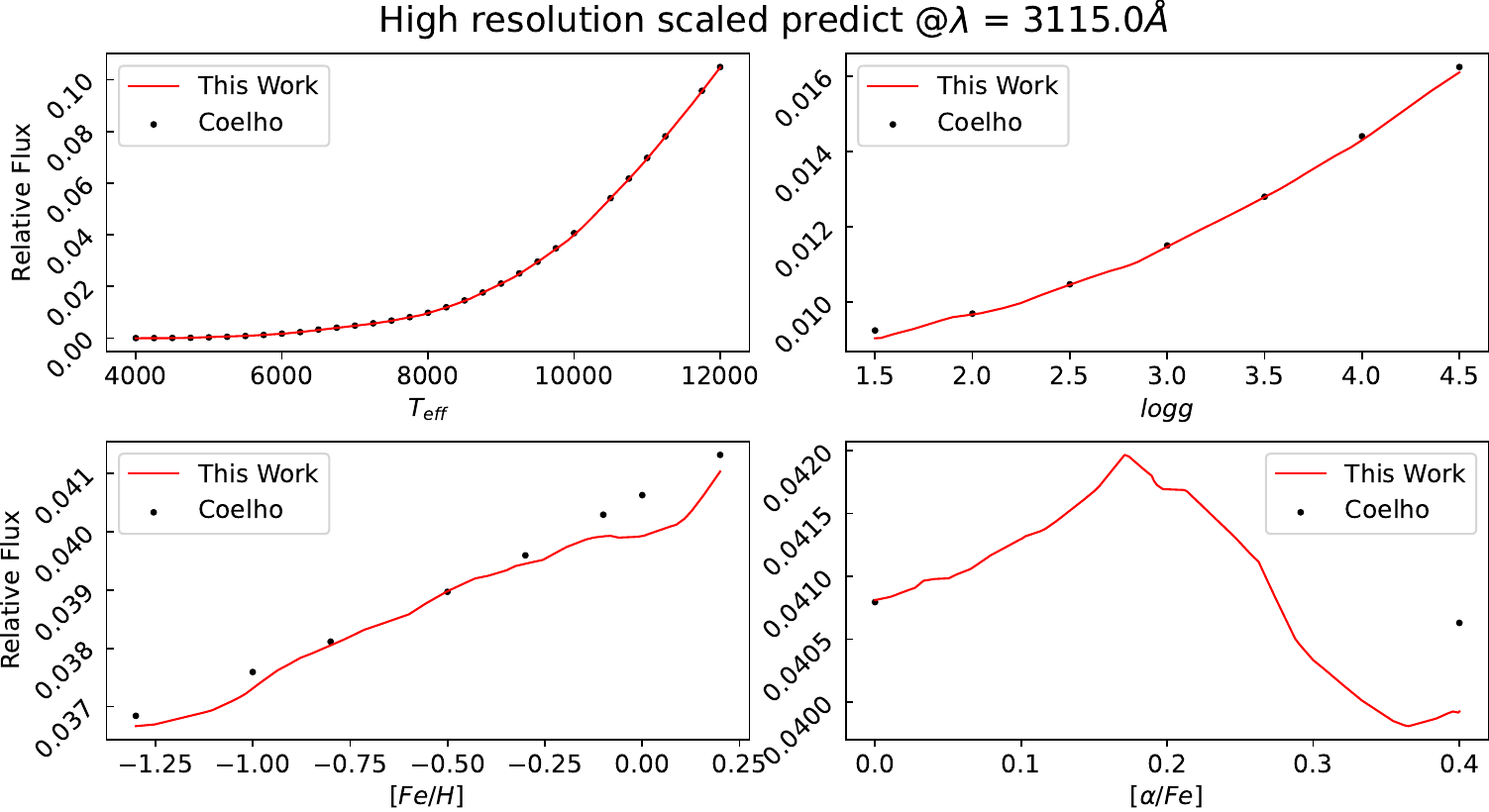}
\end{minipage}
\begin{minipage}[c]{0.5\textwidth}
    \centering
    \includegraphics[width=\textwidth]{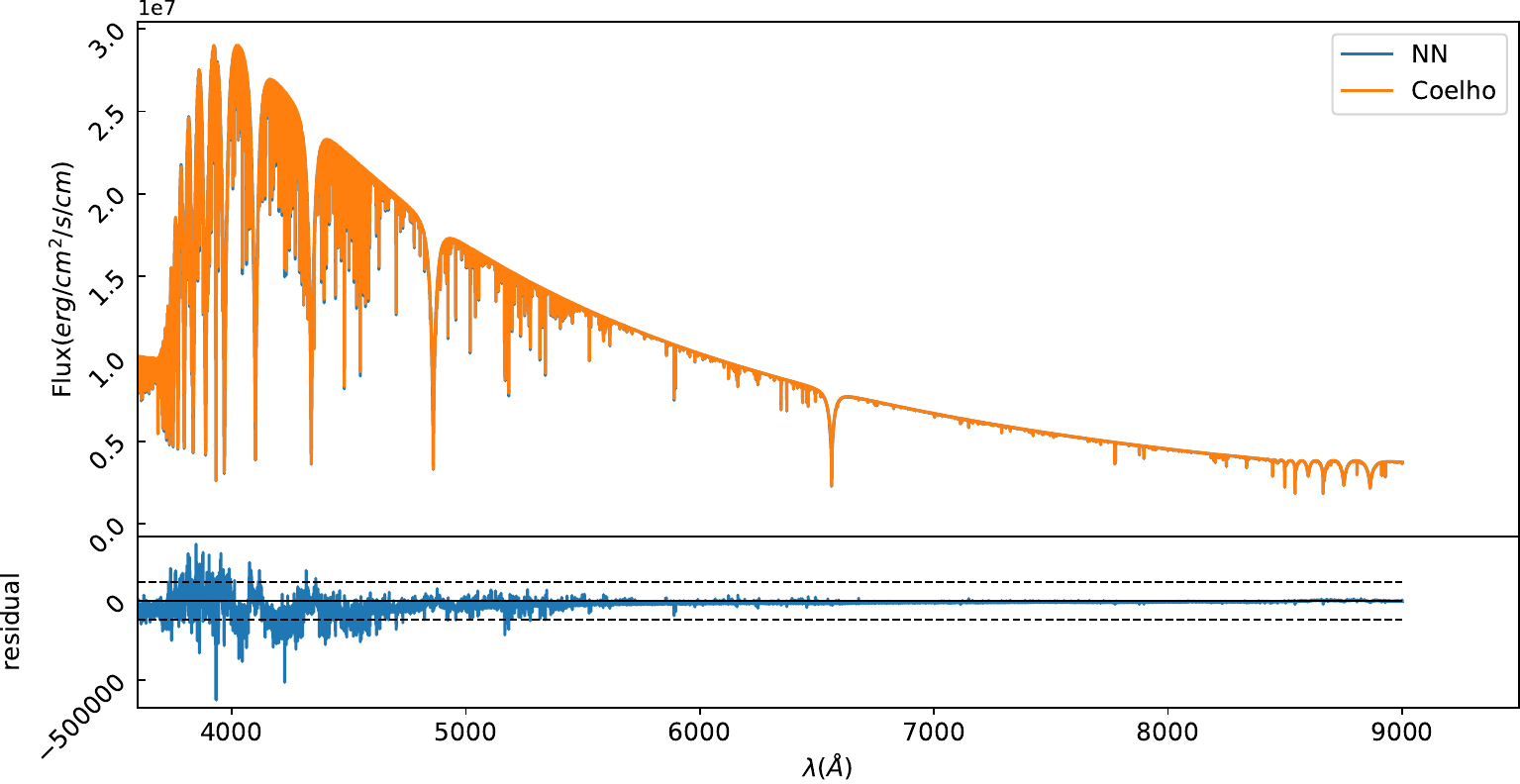}
\end{minipage}

\caption{Left panel:  Flux comparison between predicted (red curves) and theoretical values (black dots), at a random wavelength pixel for a typical MS star.
Right panel: Comparison of predicted (blue) and theoretical spectra (orange). The residual spectrum is shown in the bottom right, where the $\pm 3 \sigma$ lines are indicated by the dotted lines.   See the text for more details.}
\label{nnms}
\end{figure}

\begin{figure}[!h]
\begin{minipage}[c]{0.5\textwidth}
    \centering
    \includegraphics[width=\textwidth]{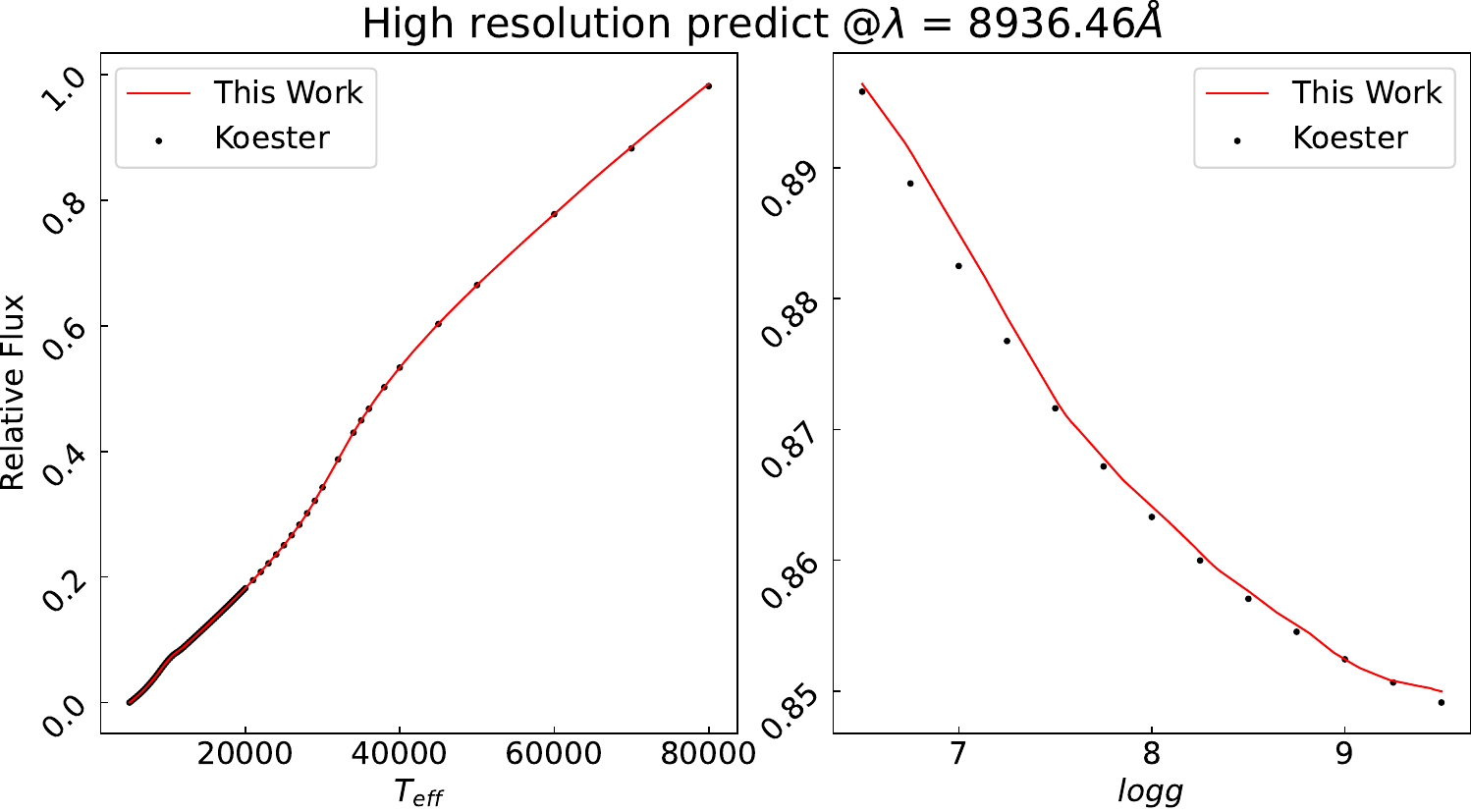}
\end{minipage}
\begin{minipage}[c]{0.5\textwidth}
    \centering
    \includegraphics[width=\textwidth]{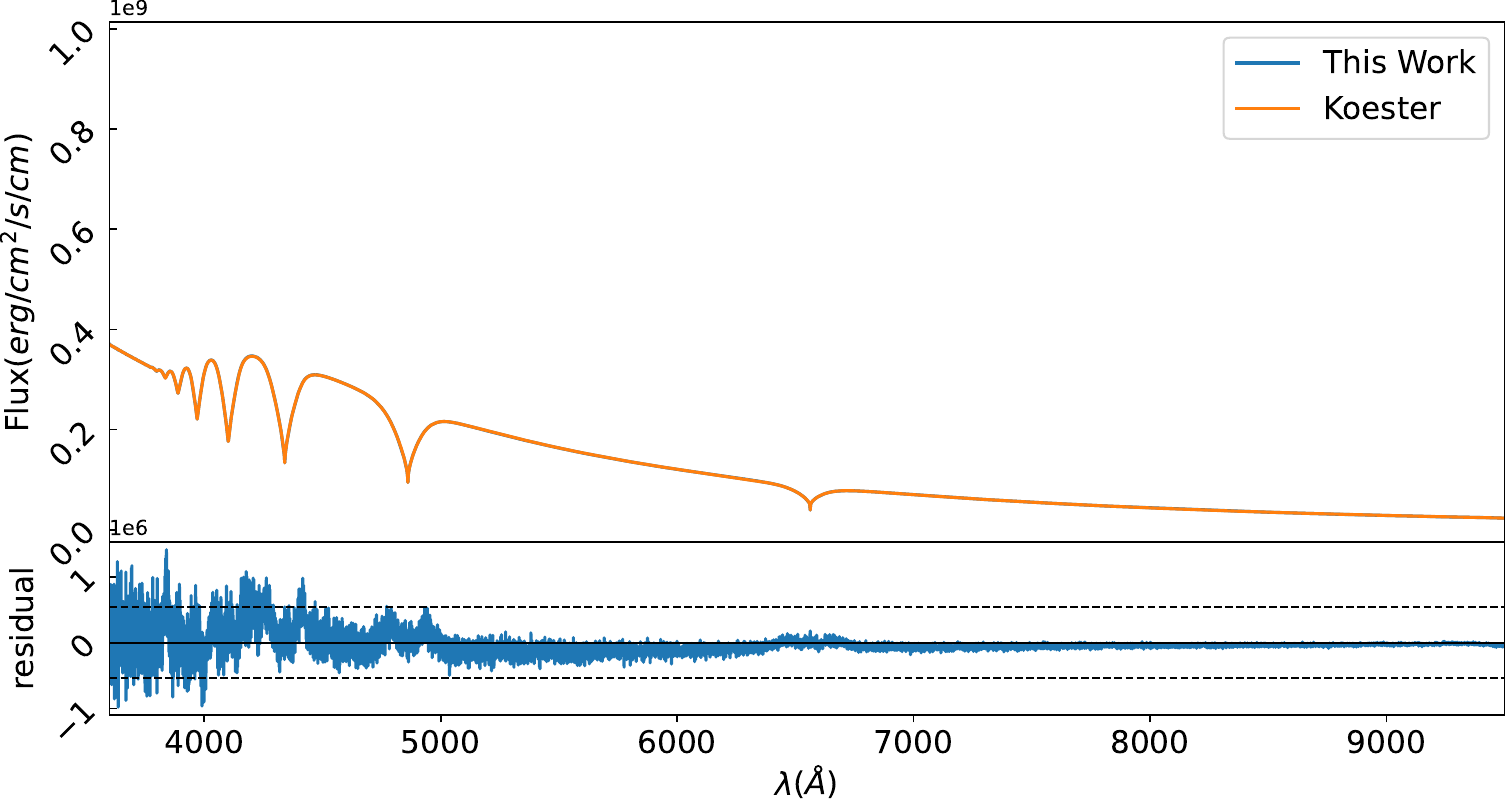}
\end{minipage}
\caption{ Flux comparison between DA prediction and \cite{koester2010white} templates. Symbols are the same as Fig. \ref{nnms}.}
\label{nnda}
\end{figure}

\begin{figure}[!h]
\begin{minipage}[c]{0.5\textwidth}
    \centering
    \includegraphics[width=\textwidth]{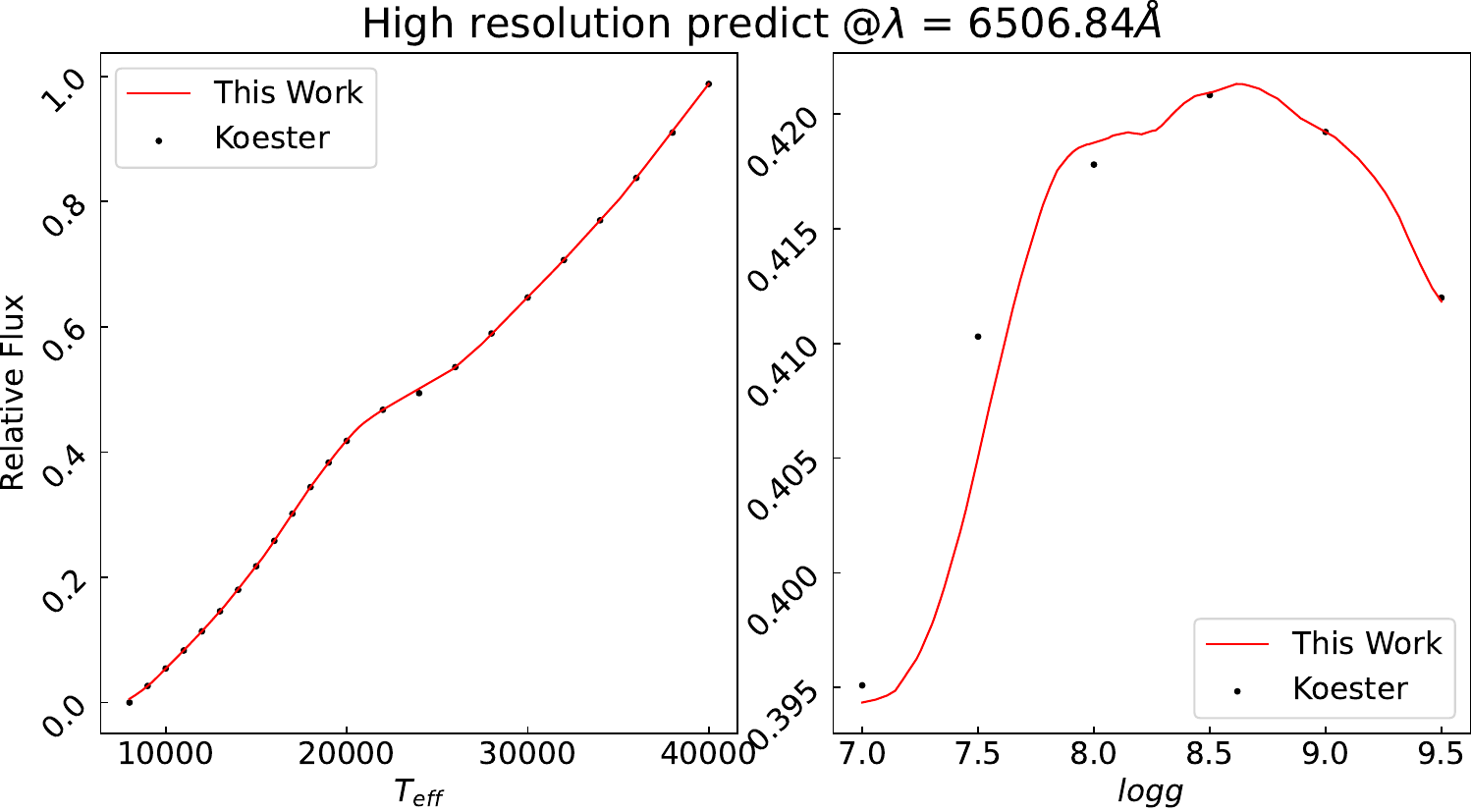}
\end{minipage}
\begin{minipage}[c]{0.5\textwidth}
    \centering
    \includegraphics[width=\textwidth]{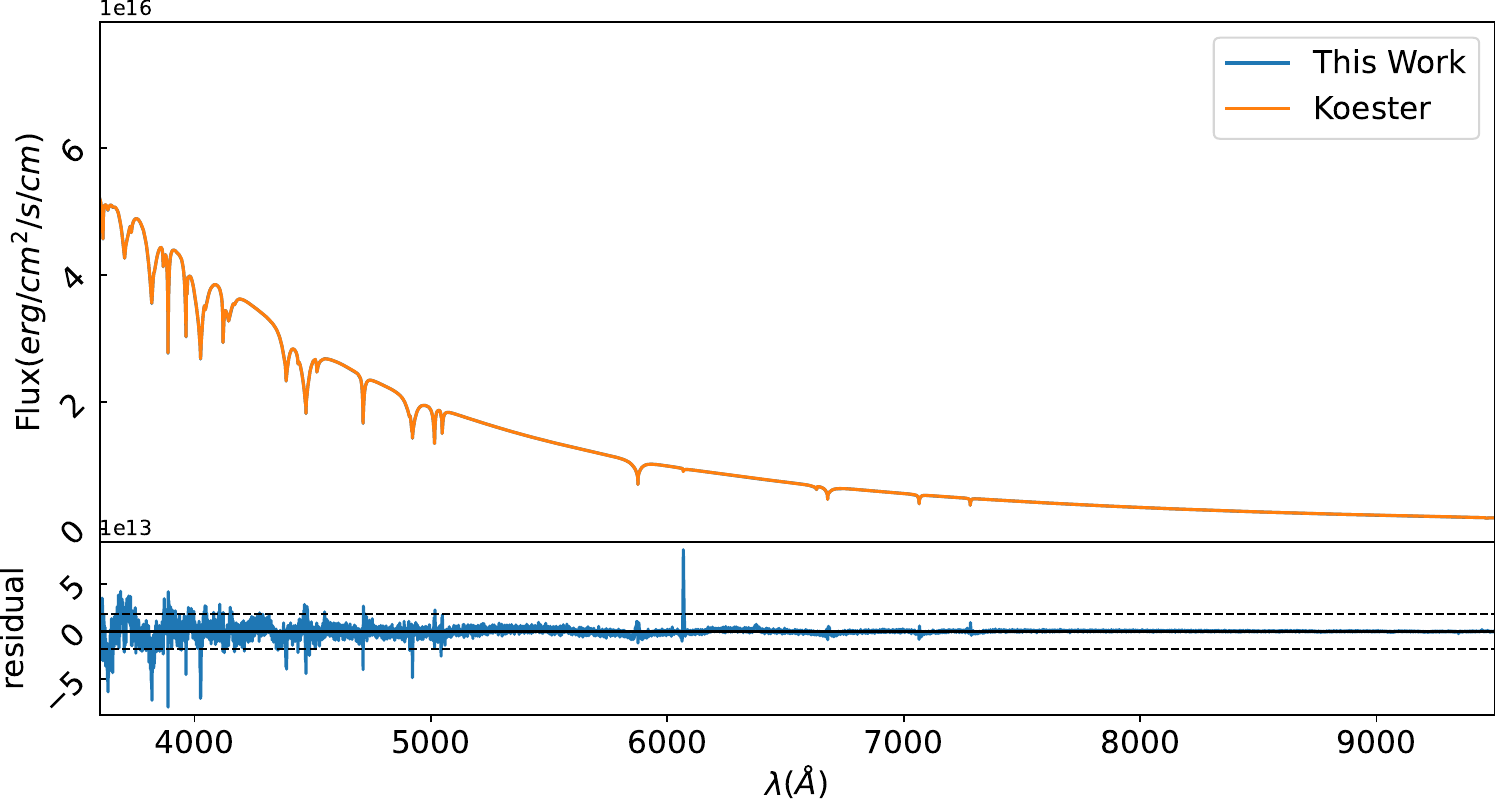}
\end{minipage}
\caption{Flux comparison between DB prediction and \cite{koester2010white} template. Symbols are the same as Fig. \ref{nnms}.}
\label{nndb}
\end{figure}

To test whether our ANN spectrum generator performs smoothly across different parameter space, we randomly select a set of parameters as inputs for the spectrum generator. The resulting flux-calibrated spectrum is compared with theoretical spectrum in the spectral library with the same parameters. The residual standard deviation is defined as 
\begin{equation}
    \label{std}
    \sigma = \sqrt{\frac{1}{N_{pix}}\sum_{j=1}^{N_{pix}}(R_{j}-\bar{R})^2},
\end{equation}

\noindent  where $R_{j}$ represents the flux residual of the $j$-th pixel, and $\bar{R}$ represents the average flux residual. The residual standard deviation only demonstrate the deviation along wavelength axis, but our spectrum generator actually works on a high-dimensional parameter space. Therefore, to test the behavior along the label axis, we randomly select one single wavelength pixel, and vary the value of a given parameter while keeping the others fixed. The new generated flux is then compared with the theoretical flux. The interpolated curves of the neural network's prediction for MS stars (the left panel of Fig. \ref{nnms}) are smooth across all stellar labels, except $[\alpha/Fe]$\footnote{There are only two model grid points for $\mathrm{[\alpha/Fe]}$, making it impossible to evaluate the accuracy of the models. Therefore, $\mathrm{[\alpha/Fe]}$ is fixed at 0 in the subsequent fitting.}. This implies that our spectrum generator does not produce overfitting artifacts.
Similar examinations are also performed on WDs (Figs. \ref{nnda} and \ref{nndb}), but the parameters are only effective temperature and surface gravity, and the number of output wavelength pixels is kept consistent with the number of the corresponding theoretical spectral wavelength pixels. 

\begin{figure}[!h]
    \centering
    \includegraphics[width=0.5\textwidth]{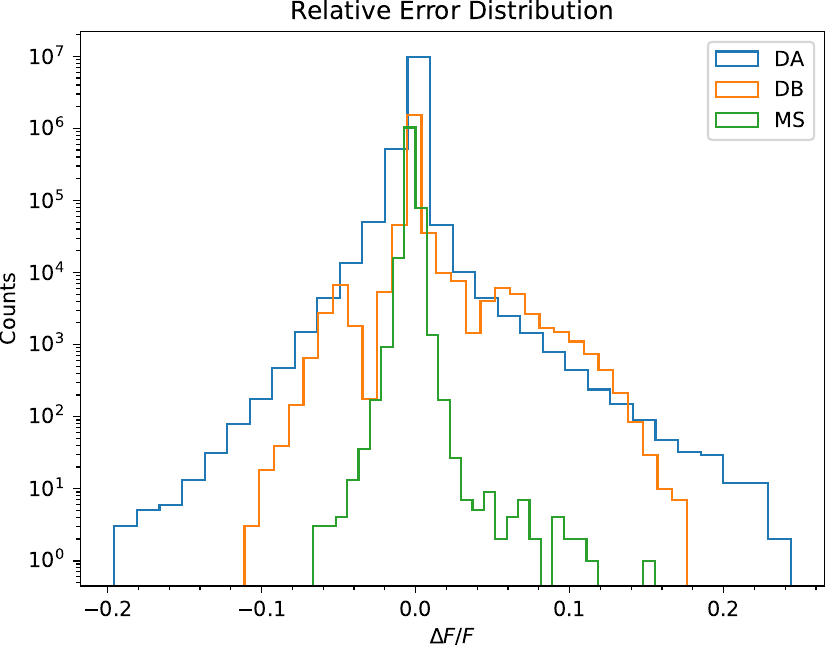}
    \caption{Histograms of the relative errors of the pixels in DA (blue), DB (orange) and MS (green) synthetic spectra. Only wavelength pixels within $3600$~\AA $<\lambda<9000 $~\AA~are considered. 
    }
    \label{errdis}
\end{figure}

Next, we examine the relative flux error of each pixel for each type of stars. Fig.\ref{errdis} shows 
the histogram for the relative flux errors for 101 MS spectra (1036 DA, and 138  DB). The mean values are closed to 0, and the relative errors are mostly less than 10\%.

\begin{figure*}[!h]
    \centering
    \includegraphics[width=0.9\textwidth]{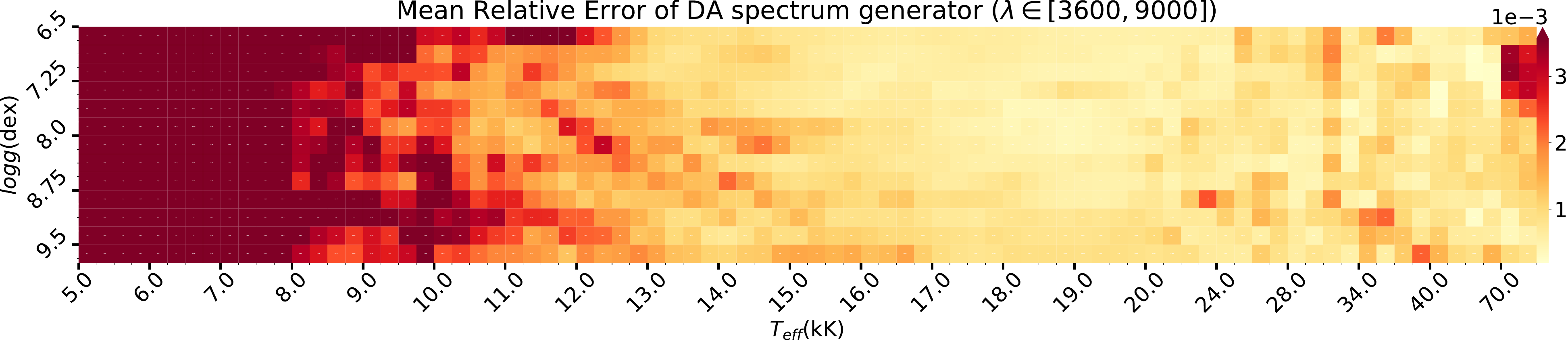}
    \caption{Mean relative errors of the entire DA datasets. For each set of parameters on the grid, we calculate the mean relative error of the corresponding spectra. These relative errors are indicated by a color bar on the right. }
    \label{daerr}
\end{figure*}

\begin{figure*}[!h]
    \centering
    \includegraphics[width=0.9\textwidth]{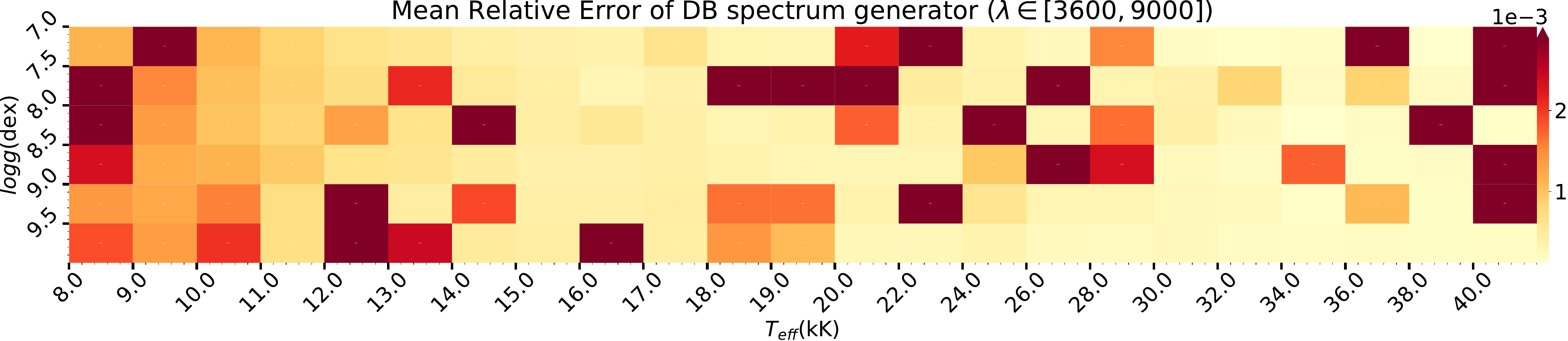}
    \caption{Same as Fig. \ref{daerr}, but for DBs.}
    \label{dberr}
\end{figure*}

\begin{figure*}[!h]
    \centering
    \includegraphics[width=0.9\textwidth]{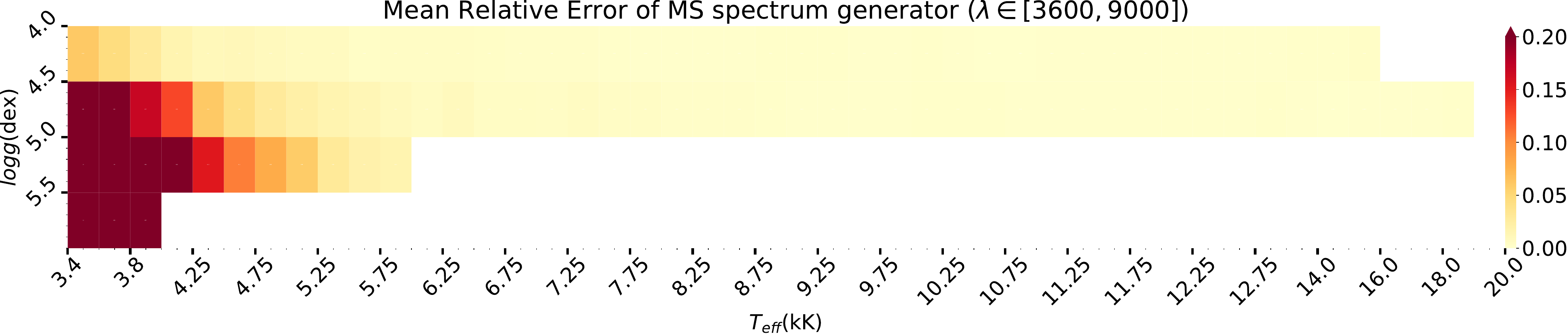}
    \caption{Same as Fig. \ref{daerr}, but for MS stars. Here we fix other two parameters: $\mathrm{[Fe/H]}=0$ and $\mathrm{[\alpha/Fe]}=0$. Moreover, the neural network doesn't work when MS stars have low temperature and high surface gravity.}
    \label{mserr}
\end{figure*}

Finally, it is also important to know which part of the parameter space (T vs. $\log g$) is more reliable.
Figures \ref{daerr}, \ref{dberr}, and \ref{mserr} show maps of relative errors for DA, DB, and MS stars, respectively. We note that (1) The relative errors reach  $\sim 0.3\%$ for cool pure-H white dwarf model ($\mathrm{T}<8,000 \mathrm{K}$); (2) The relative errors of pure-He white dwarf models are quite uniform across most parameter space (no larger than $0.1\%$); (3) The relative errors MS stars are small (<5\%) for hotter stars ($\mathrm{T_{MS}}>4,250 \mathrm{K}$), and increase to  $\sim 10\% $ for cooler stars.



\newpage
\section{Fitting result of 14 white dwarf binary candidates}
\begin{table}[!h]
    \centering
    \caption{Fitted stellar parameters.}
\label{var1}
\begin{tabular}{cccccccc}
\hline
Name&Type&$\mathrm{T_{eff}(K)}$&$logg (cgs)$&[Fe/H]&$\mathrm{M}$ $(\mathrm{M_{\odot}})$&$\mathrm{R}$ $(10^{-2}\mathrm{R_{\odot}})$&LC type\\\hline
\multirow{2}{*}{ZTF J110045.15+521043.80}&DA&$28468^{+567}_{-609}$&$7.658^{+0.028}_{-0.03}$&$/^{+/}_{-/}$&$0.482^{+0.012}_{-0.013}$&$1.705^{+0.05}_{-0.055}$&\multirow{2}{*}{EA}\\
&/&$/^{+/}_{-/}$&$/^{+/}_{-/}$&$/^{+/}_{-/}$&$/^{+/}_{-/}$&$/^{+/}_{-/}$&\\\hline
\multirow{2}{*}{ZTF J140847.17+295044.87}&DA&$31118^{+856}_{-905}$&$8.058^{+0.042}_{-0.044}$&$/^{+/}_{-/}$&$0.681^{+0.025}_{-0.026}$&$1.278^{+0.064}_{-0.06}$&\multirow{2}{*}{EA}\\
&MS&$3620^{+57}_{-49}$&$5.14^{+0.02}_{-0.01}$&$0.13^{+0.05}_{-0.05}$&$0.1^{+0.02}_{-0.02}$&$14.44^{+0.63}_{-0.67}$&\\\hline
\multirow{2}{*}{ZTF J082145.29+455923.46}&DA&$78106^{+2427}_{-4643}$&$7.602^{+0.019}_{-0.037}$&$/^{+/}_{-/}$&$0.596^{+0.012}_{-0.024}$&$2.022^{+0.083}_{-0.043}$&\multirow{2}{*}{EA}\\
&MS&$3352.0^{+160}_{-261}$&$4.68^{+0.08}_{-0.05}$&$0.2^{+0.29}_{-0.26}$&$0.3^{+0.15}_{-0.08}$&$41.57^{+2.55}_{-3.94}$&\\\hline
\multirow{2}{*}{ZTF J034137.67+011027.88}&MS&$5768^{+22}_{-22}$&$4.719^{+0.019}_{-0.008}$&$-1.132^{+0.058}_{-0.022}$&$0.707^{+0.027}_{-0.012}$&$60.844^{+0.485}_{-0.489}$&\multirow{2}{*}{EB}\\
&/&$/^{+/}_{-/}$&$/^{+/}_{-/}$&$/^{+/}_{-/}$&$/^{+/}_{-/}$&$/^{+/}_{-/}$&\\\hline
\multirow{2}{*}{ZTF J143257.08+491143.01}&MS&$8620^{+305}_{-366}$&$4.646^{+0.163}_{-0.176}$&$-0.981^{+0.106}_{-0.118}$&$1.327^{+0.097}_{-0.094}$&$90.681^{+17.337}_{-15.049}$&\multirow{2}{*}{EB}\\
&MS&$5774^{+621}_{-982}$&$4.3^{+0.17}_{-0.12}$&$-1.27^{+0.36}_{-0.33}$&$0.61^{+0.11}_{-0.19}$&$91.25^{+12.65}_{-17.61}$&\\\hline
\multirow{2}{*}{ZTF J093053.12+192430.20}&MS&$6136^{+304}_{-290}$&$4.339^{+0.053}_{-0.049}$&$-1.294^{+0.072}_{-0.044}$&$0.714^{+0.105}_{-0.095}$&$94.733^{+5.501}_{-5.865}$&\multirow{2}{*}{ELL}\\
&MS&$8076^{+105}_{-131}$&$4.8^{+0.07}_{-0.08}$&$-1.13^{+0.1}_{-0.12}$&$1.09^{+0.05}_{-0.06}$&$68.58^{+5.01}_{-4.58}$&\\\hline
\multirow{2}{*}{ZTF J140118.80-081723.51}&DA&$22762^{+978}_{-4679}$&$6.912^{+0.012}_{-0.549}$&$/^{+/}_{-/}$&$0.273^{+0.007}_{-0.014}$&$3.031^{+2.106}_{-0.043}$&\multirow{2}{*}{ELL}\\
&MS&$7320^{+210}_{-7}$&$5.27^{+0.01}_{-0.01}$&$-1.3^{+0.0}_{-0.0}$&$0.65^{+0.05}_{-0.01}$&$30.77^{+0.17}_{-0.21}$&\\\hline
\multirow{2}{*}{ZTF J100414.83+381833.96}&DA&$74513^{+9192}_{-11270}$&$7.371^{+0.052}_{-0.063}$&$/^{+/}_{-/}$&$0.529^{+0.027}_{-0.034}$&$2.486^{+0.105}_{-0.094}$&\multirow{2}{*}{ELL}\\
&MS&$5175^{+28}_{-25}$&$4.17^{+0.0}_{-0.0}$&$-1.19^{+0.04}_{-0.05}$&$0.68^{+0.01}_{-0.01}$&$111.74^{+0.94}_{-1.01}$&\\\hline
\multirow{2}{*}{ZTF J151712.07+541937.82}&MS&$4833^{+186}_{-172}$&$4.565^{+0.046}_{-0.037}$&$-0.934^{+0.101}_{-0.149}$&$0.527^{+0.077}_{-0.055}$&$62.737^{+3.978}_{-4.79}$&\multirow{2}{*}{ELL}\\
&MS&$5896^{+222}_{-193}$&$4.37^{+0.19}_{-0.21}$&$-1.15^{+0.11}_{-0.25}$&$0.67^{+0.08}_{-0.16}$&$88.07^{+14.58}_{-12.62}$&\\\hline
\multirow{2}{*}{ZTF J162454.48+494903.46}&MS&$4834^{+41}_{-42}$&$4.431^{+0.017}_{-0.008}$&$-1.076^{+0.186}_{-0.118}$&$0.398^{+0.014}_{-0.008}$&$63.591^{+0.653}_{-1.392}$&\multirow{2}{*}{ELL}\\
&DB&$23986^{+212}_{-482}$&$7.47^{+0.15}_{-0.16}$&$/^{+/}_{-/}$&$0.37^{+0.07}_{-0.07}$&$1.87^{+0.23}_{-0.2}$&\\\hline
\multirow{2}{*}{ZTF J173005.10+430450.20}&MS&$7473^{+45}_{-35}$&$4.408^{+0.014}_{-0.009}$&$-1.3^{+0.084}_{-0.049}$&$1.189^{+0.019}_{-0.014}$&$112.885^{+0.902}_{-1.121}$&\multirow{2}{*}{ELL}\\
&/&$/^{+/}_{-/}$&$/^{+/}_{-/}$&$/^{+/}_{-/}$&$/^{+/}_{-/}$&$/^{+/}_{-/}$&\\\hline
\multirow{2}{*}{ZTF J074449.45+290709.64}&MS&$8877^{+104}_{-111}$&$4.805^{+0.039}_{-0.051}$&$-1.181^{+0.14}_{-0.166}$&$1.189^{+0.014}_{-0.077}$&$71.471^{+4.202}_{-3.173}$&\multirow{2}{*}{ELL}\\
&MS&$6067^{+190}_{-217}$&$4.5^{+0.07}_{-0.07}$&$-1.3^{+0.1}_{-0.06}$&$0.72^{+0.06}_{-0.08}$&$79.17^{+8.5}_{-8.24}$&\\\hline
\multirow{2}{*}{ZTF J151411.35+444000.23}&MS&$6761^{+382}_{-558}$&$4.894^{+0.301}_{-0.073}$&$-0.2^{+0.478}_{-0.139}$&$1.125^{+0.087}_{-0.127}$&$62.773^{+7.294}_{-28.178}$&\multirow{2}{*}{ELL}\\
&MS&$7862^{+167}_{-511}$&$4.55^{+0.15}_{-0.21}$&$-1.15^{+0.16}_{-0.16}$&$1.17^{+0.09}_{-0.09}$&$95.25^{+19.06}_{-12.53}$&\\\hline
\multirow{2}{*}{ZTF J093934.58+531751.45}&DA&$39796^{+1771}_{-3673}$&$7.625^{+0.125}_{-0.124}$&$/^{+/}_{-/}$&$0.5^{+0.058}_{-0.056}$&$1.805^{+0.201}_{-0.194}$&\multirow{2}{*}{EW}\\
&MS&$5503^{+67}_{-66}$&$4.72^{+0.04}_{-0.04}$&$-1.07^{+0.09}_{-0.1}$&$0.7^{+0.03}_{-0.03}$&$59.96^{+3.07}_{-3.21}$&\\\hline
\end{tabular}
\tablefoot{ Column 1 shows the ZTF IDs, Column 2 shows the best-fit components. Column 3-7 are the corresponding atmospheric parameters ($\mathrm{T_{eff}}, \mathrm{logg}, \mathrm{[Fe/H]}$ for MS stars and $\mathrm{T_{eff}}, \mathrm{logg}$ for DA,DB stars), masses and radius. Column 8 are the binary types classified using their light curves \citep{Ren_2023}. 
The label "$/$" means no value; see text for more details.}
\end{table}

\clearpage

\begin{figure}
    \centering
    \includegraphics[width=\textwidth]{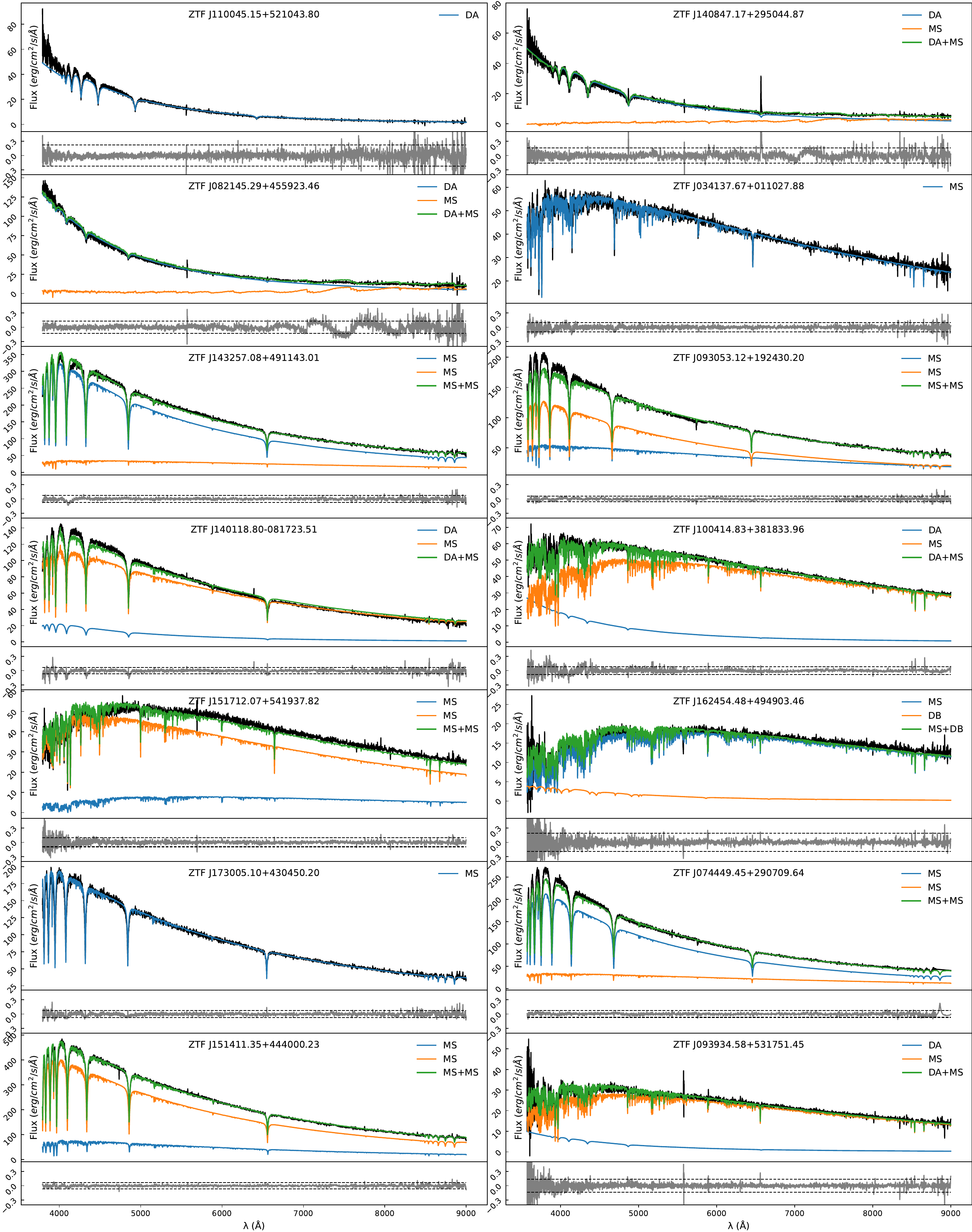}
    \caption{Fitting results of 14 CWDB candidates. Symbols are the same as in Fig. \ref{sampledwds}.}
    \label{ff}
\end{figure}

\end{appendix}
\end{document}